# Entrepreneurship, Institutions, and Economic Growth: Does the Level of Development Matter?

**Christopher Boudreaux**
Department of Economics
Florida Atlantic University
777 Glades Road, KH 145
Boca Raton, FL 33431 USA
Tel: 1-561-297-3221
Email: cboudreaux@fau.edu

**ABSTRACT**

Entrepreneurship is often touted for its ability to generate economic growth. Through the creative-destructive process, entrepreneurs are often able to innovate and outperform incumbent organizations, all of which is supposed to lead to higher employment and economic growth. Although some empirical evidence supports this logic, it has also been the subject of recent criticisms. Specifically, entrepreneurship does not lead to growth in developing countries—only in more developed countries with higher income levels. Using Global Entrepreneurship Monitor (GEM) data for a panel of 83 countries from 2002 to 2014, we examine entrepreneurship's contribution towards economic growth. Our evidence validates earlier studies' findings but also exposes previously undiscovered findings. That is, we find that entrepreneurship encourages economic growth but not in developing countries. In addition, our evidence finds that a country's institutional environment—measured by GEM's Entrepreneurial Framework Conditions (EFCs), only contributes to economic growth in more developed countries but not in developing countries. These findings have important policy implications. Namely, our evidence contradicts policy proposals that suggest entrepreneurship and the adoption of pro-market institutions that support it to encourage economic growth in developing countries. Our evidence suggests these policy proposals will be unlikely to generate the economic growth desired.

1. **Introduction**

It is often suggested that entrepreneurship is valuable because of its ability to generate economic growth and development (Acs, 2006; Acs et al., 2018; Acs & Szerb, 2007; Audretsch et al., 2006; Baumol, 1986; Baumol & Strom, 2007; Bosma et al., 2018; Braunerhjelm et al., 2010; Schumpeter, 1934; Wennekers & Thurik, 1999). Endogenous growth theory (Lucas, 1988; Romer, 1986, 1990), for instance, posits that economic growth depends on knowledge accumulation and its diffusion through both incumbents and entrepreneurial activities (Braunerhjelm et al., 2010). Investments in human capital and R&D create knowledge for incumbents but also create knowledge spillovers for new entrepreneurs (Acs et al., 2009; Audretsch & Keilbach, 2007; Braunerhjelm et al, 2018).

It should be no surprise then to find an abundance of claims like "entrepreneurship is the main vehicle of economic development" (Anokhin, Grichnik, & Hisrich, 2008, p. 117), "the more entrepreneurs there are in an economy the faster it will grow" (Dejardin, 2000, p. 2), and "the engine of economic growth is the entrepreneur" (Holcombe, 1998, p. 60). Thus, it is often taken for granted that entrepreneurship encourages economic development (Naudé, 2009). Despite these claims, however, there is now evidence to suggest that the relationship between entrepreneurship and economic growth does not hold for developing countries (Sautet, 2013) and might even be negative (Van Stel et al., 2005). The reality is, "We actually know very little about whether and how entrepreneurship either contributes or does not contribute to economic growth in developing countries" (Autio, 2008, p. 2).

The purpose of our study is to revisit the policy claim that entrepreneurship unequivocally encourages economic growth. Some evidence suggests that both too much and too little entrepreneurship detracts from long-run country growth rates (Carree et al., 2002). More importantly, if entrepreneurship only facilitates economic growth in developed countries and has no effect in developing countries (Sautet, 2013; Van Stel et al., 2005), then scholars and policy makers should reconsider how entrepreneurship policy recommendations (Mason & Brown, 2013; Shane, 2009) might fail to extend to other contexts. Using Global Entrepreneurship Monitor (GEM) data for a panel of 83 countries from 2002 to 2014, we examine entrepreneurship's contribution towards economic growth. We estimate a mixture model to test the hypothesis that two regimes exist in the data—that entrepreneurship encourages growth for one group (i.e., developed countries) and does not encourage growth for another group (i.e., developing countries). Our evidence supports this hypothesis and uncovers other important findings. Specifically, we find that a country's institutional environment—measured by GEM's Entrepreneurial Framework Conditions (EFCs)—only contributes to economic growth in more developed countries but not in developing countries, which is an additional important finding.

Our findings make several contributions to the literature. First, our study makes an important update to the literature on entrepreneurship, economic growth, and economic development (Acs et al., 2008; Naudé, 2009; Sautet, 2013; Urbano et al., 2018; Van Stel et al., 2005). Specifically, our study most closely resembles the study, "*The Effect of Entrepreneurial Activity on National Economic Growth*" (Van Stel et al., 2005). In this study, Van Stel and his colleagues discover that total early-stage entrepreneurial activity (TEA) encourages economic growth in high-income countries but discourages growth in low-income countries. Their study, while undoubtedly important, makes no distinction between the different *types* of entrepreneurship. Recent insights, for instance, suggest that OME is more likely to lead to economic growth than NME (Hessels, Gelderen, & Thurik, 2008; Nikolaev, Boudreaux, & Palich, 2018). We use this insight to suggest that the reason that entrepreneurship encourages economic development in developed countries and discourages it in developing countries is because of their different levels of OME and NME. Furthermore, Van Stel et al. (2005) examine the relationship between TEA and economic growth using only a cross-section of 36 countries. Thus while a good start, it fails to account for important differences between countries, it uses a small sample at only one point in time, and it does not include other relevant explanatory variables that might influence economic growth, which potentially introduces omitted variable bias. We therefore revisit their research questions. Our evidence supports their original findings and extends their analysis to the relative contributions of OME and NME. Our findings are also consistent with more recent theoretical contributions on the failure of entrepreneurship to encourage economic growth in developing countries (Sautet, 2013). Because we find that OME encourages economic growth in high-income countries and NME discourages economic growth in low-income countries, our findings imply that policymakers might look to reduce NME in low-income countries to increase economic growth, which has been a previously overlooked aspect of the relationship.

Second, these findings have important policy implications. We find that neither entrepreneurship nor institutional conditions encourage economic growth in developing countries. This suggests that policies designed to encourage entrepreneurship in developed countries (Acs et al., 2016; Mason & Brown, 2013; Shane, 2009) might be unlikely to be successful in the developing world. Recent contributions, for example, argue that pro-market institutions encourage entrepreneurship, which in turn, contributes to economic growth (Bjørnskov & Foss, 2016; Bosma et al., 2018; Bradley & Klein, 2016). Yet there has been little attention given to these relationships in the developing world (Naudé, 2009). We therefore believe that entrepreneurship policy in the developing world has largely been overlooked and deserves additional attention, especially because entrepreneurship policies have arguably more importance for growth in developing countries where entrepreneurship helps to alleviate poverty (Alvarez & Barney, 2014; Bruton et al., 2013; Court & Maxwell, 2005). Our study builds on this literature by highlighting one potential conduit to increase economic growth in developing countries—by reducing the prevalence of necessity entrepreneurship.

Third, we find that institutions are important antecedents of economic growth but only in developed countries. This finding supports earlier studies on entrepreneurship, institutions, and economic growth (Acs et al., 2008; Van Stel et al., 2005). Yet, it contradicts generic statements that imply that institutions unequivocally encourage economic growth (Acemoglu et al., 2005; Bosma et al., 2018; Dawson, 1998; Dollar & Kraay, 2003).

Lastly, we synthesize recent theoretical and empirical developments to explain the mechanisms behind entrepreneurship and economic growth. We explore how institutional theorists use Coleman's bathtub model to explain the pathway from institutions to entrepreneurship to economic growth (Bjørnskov & Foss, 2016; Bradley & Klein, 2016; Coleman, 1990; Kim, Wennberg, & Croidieu, 2016). We also explore the role that knowledge serves in the spillover theory of entrepreneurship of endogenous growth theory (Acs et al., 2009; Acs et al., 2012; Audretsch & Keilbach, 2007; Braunerhjelm et al., 2010), and we examine how this relates to the different effects of entrepreneurship on economic growth across Porter's stages of competitiveness (Porter, 1990).

## 2. Models and Literature Development

There have been many contributions to the literature on entrepreneurship and economic growth in recent years (see e.g., Urbano, Aparicio, & Audretsch, (2018) for a recent review). Although these contributions share much in common, we can separate their contributions based on three different explanations for why some countries have found entrepreneurship to be a better predictor of growth than others. This section reviews these strands of the literature to gain insights towards how entrepreneurship, institutions, and policy all affect economic growth.

### 2.1. The nexus of institutions, entrepreneurship, and growth

The first explanation focuses on the role that institutions serve in the relationship between entrepreneurship and growth (Acs et al., 2008; Acs et al., 2017; Acs et al., 2018; Bjørnskov & Foss, 2013, 2016). This strand argues that pro-market institutions encourage productive entrepreneurship and discourage unproductive entrepreneurship (Baumol, 1990; Sobel, 2008), which translates into greater economic growth (Bosma et al., 2018). In a complete model, pro-market institutions lead to higher rates of entrepreneurial entry (Urbano & Alvarez, 2014) and higher rates of entrepreneurial entry lead to more economic growth (Braunerhjelm et al., 2010). Pro-market institutions can encourage a protection of property rights, which are important for capital accumulation and entrepreneurial investment (De Soto, 2000), but they can also reduce the adverse effects of regulation on entrepreneurial entry (Djankov et al., 2002; Ho & Wong, 2007; Klapper et al., 2006; Van Stel et al., 2007). Recent contributions have modeled this mechanism using a multi-stage analysis where institutions affect entrepreneurship in the first stage, which subsequently affects economic growth in the second stage (Bosma et al., 2018; Urbano et al., 2018). One conceptual way to model this relationship is through the Coleman bathtub model (Bjørnskov & Foss, 2016; Bradley & Klein, 2016; Kim et al., 2016), which is illustrated in Figure 1.

**[Insert Figure 1]**

In *Foundations of Social Theory,* Coleman (1990) uses the bathtub model to illustrate how macro-level structures affect micro-level behaviors and actions. Entrepreneurship scholars have extended this model to examine how institutional conditions contribute to economic growth through the operational channel of entrepreneurship (Bjørnskov & Foss, 2016; Bradley & Klein, 2016; Kim et al., 2016). First, institutions emerge at the macro-level. Institutions define the rules of the game (North, 1990). They can be regulative, normative, or cultural cognitive (Scott, 1995), and they determine economic behavior. Institutional conditions, when applied to entrepreneurship, encourage productive entrepreneurship (Bjørnskov & Foss, 2008; Boudreaux, 2014; Boudreaux et al., 2018; Bowen & Clercq, 2008; McMullen et al., 2008; Nikolaev et al., 2018; Nyström, 2008) and discourage unproductive or destructive entrepreneurship (Baumol, 1990; Boudreaux, et al., 2018; Sobel, 2008). GEM describes these pro-market institutions as the entrepreneurial framework conditions (EFCs) that encourage or hinder entrepreneurship activity. Institutional conditions determine the micro-level behavior of entrepreneurs by encouraging entrepreneurial traits and decision making (path B in Figure 1). These entrepreneurial traits and decisions such as opportunity recognition, entrepreneurial self-efficacy, a lack of fear of failure, and social capital, in turn, affect entrepreneurial entry and participation, which is a robust finding in the literature on the cognitive traits behind entrepreneurship (Boudreaux & Nikolaev, 2018; Boudreaux et al., 2017; De Clercq et al., 2013) This is illustrated by path C in Figure 1. Finally, in the aggregate, entrepreneurial entry and participation affect economic growth, which is reported at the macro-level and reported as path D in Figure 1. Thus, rather than positing that entrepreneurship affects economic growth merely at the macro-level (i.e., path A in Figure 1) as earlier cross-country studies suggested (Bjørnskov & Foss, 2008; Nyström, 2008), the Coleman bathtub model provides the insight that micro-foundations help explain how institutions encourage economic growth—through the channel of entrepreneurship.

2.2. Levels of development and economic growth

The second explanation argues that entrepreneurship can encourage growth, but it is the *entrepreneurship type* that matters (Ács & Varga, 2005). This explanation argues that the level of economic development determines the ability of entrepreneurship to contribute to economic growth (Sautet, 2013; Van Stel et al., 2005). Porter (1990) defines competitiveness according to three stages: (1) *factor-driven stage*, (2) *efficiency-driven stage,* and (3) *innovation-driven stage* (Acs et al., 2008). *Factor-driven* economies are dominated by the production of commodities and low value-added products. In this stage, high rates of non-agricultural self-employment are prevalent. Importantly, *factor-driven* economies do not create knowledge or innovation, which suggests limited effects on economic growth (Acs et al., 2008). Countries begin in the *factor-driven stage* but transition into the *efficiency-driven stage.* In this second stage, countries focus predominately on efficiency in production and a highly educated workforce, which are necessary to adapt to technological developments and to exploit economies of scale (Acs et al., 2008). Importantly, during this second stage, there is a transition from self-employment to wage-employment because of the substitution between capital and labor that arises during this stage. This substitution increases returns from working and lowers the returns from self-employment (Acs et al., 2008). Lastly, countries transition from the *efficiency-driven stage* to the *innovation-* *driven stage*. In this third stage, countries experience a decline in manufacturing and an increase in services, which provide more opportunities for entrepreneurship (Acs et al., 2008). In addition, improvements in information technology have also enhanced the returns to entrepreneurship (Jorgenson, 2001).

Based on these insights, we expect that developed countries, which are predominately in the *innovation-driven stage* (Acs et al., 2008), have higher rates of high-growth entrepreneurship. This linkage and the finding that innovative start-up activity leads to more economic growth than the typical entrepreneur (Mueller, 2007), suggest that entrepreneurship in developed countries is more likely to positively contribute to economic growth (Sternberg & Wennekers, 2005). Developing countries, in contrast, are in the *efficiency-driven stage* or the *factor-driven stage* (Acs et al., 2008), which are likely to have higher rates of necessity-entrepreneurship, which has limited effects on economic growth (Sternberg & Wennekers, 2005). Although some of these developing countries have transitioned away from self-employment, they often experience a corresponding reduction in opportunity entrepreneurship (Acs et al., 2008), due to the substitution towards wage-employment (Aquilina et al., 2006). Thus, we also expect the effects of opportunity entrepreneurship to be more limited in developing countries. Based on this literature review, we hypothesize the following relationships:

**Hypothesis 1:** *Entrepreneurship is positively associated with economic growth but only in developed countries and for opportunity-motivated entrepreneurship.*

**Hypothesis 2:** *Entrepreneurship is negatively associated with economic growth but only in developing countries and for necessity-motivated entrepreneurship.*

**[Insert Figure 2]**

**[Insert Figure 3]**

## 3. Data and Analysis

### 3.1. Data

We explore how entrepreneurship and institutions affect economic growth using data from 83 countries between 2002 to 2014 using Global Entrepreneurship Monitor (GEM). Our study uses data from several sources. We use data from GEM's (Reynolds et al., 2005) Adult Population Survey (APS) to examine the characteristics, motivations and ambitions of individuals starting businesses and the social attitudes towards entrepreneurship (Douglas & Shepherd, 2002; Wiklund, Davidsson, & Delmar, 2003). Using GEM's methodology, we extract our ecosystems measures from the Entrepreneurial Framework Conditions (EFCs), which propose that conditions can either enhance or hinder new business creation (GEM, 2016).

**[Insert Table 1]**

Table 1 reports the descriptive statistics for the entire sample and the correlation matrix for all variables included in the study. The average level of GDP per capita is $25,800. On average, 7.94 percent of individuals participate in OME and 2.91 percent participate in NME. Forty percent of individuals know other entrepreneurs, 41 percent are actively looking for opportunities in the next six months (i.e., opportunity recognition), 50 percent believe they have the skills and knowledge required to start a business (i.e., entrepreneurial self-efficacy), and 37 percent respond that the fear of failure might prevent them from starting a business. On average, countries have a fairly high percent of status, attention, and positive perceptions in society. Sixty-five percent of individuals respond hat entrepreneurship is a desirable career choice, 70 percent state that there is a high status for entrepreneurs, and 60 percent respond that there is media attention for entrepreneurs.

Interestingly, both OME and NME are negatively correlated with GDP per capita. Although we expected this relationship for NME, we hypothesized a positive relationship between OME and GDP per capita, at least for developed countries. One explanation for the negative correlation between OME and GDP is that developing countries usually have high rates of both NME and OME (Acs et al., 2004; Nikolaev et al., 2018). We find support for this in our data since there is positive correlation (r = 0.77) between OME and NME. We observe a positive correlation between EFCs and GDP per capita but a negative correlation between most other variables and GDP per capita. We note that the correlation matrix reflects differences *between* countries rather than changes *within* countries over time. Thus, we expect positive correlations between these variables and GDP per capita to emerge in our longitudinal analysis because we examine changes both within a country and over time.

## 3.2. Measures
### 3.2.1. Dependent Variable: Ln GDP

We measure economic growth, our dependent variable, as the natural logarithm of gross domestic product per capita (GDP). This variable is provided by the World Bank database and is measured as real GDP (i.e., adjusted for inflation) and adjusted for international comparisons (i.e., purchasing power parity (PPP)). This variable is collected for all available years for our data, which is from 2002 to 2014. We transformed this measure using the natural logarithm, which is consistent with the literature on growth (Islam, 1995) and the nexus of entrepreneurship and growth (Bosma et al., 2018).

### 3.2.2. *Entrepreneurship: Opportunity and Necessity*

Following recent work (Boudreaux, Nikolaev, & Klein, 2018), we define entrepreneurship as an "attempt at a new business or new venture creation, such as self-employment, a new business organization, or the expansion of an existing business" (GEM, 2016). We gather entrepreneurship data from Global Entrepreneurship Monitor (GEM)'s Adult Population Survey (APS). The APS measures the level and nature of entrepreneurial activity around the world, and it is administered by GEM National Teams to survey a representative national sample of a minimum of 2000 respondents for each survey. GEM teams conduct these surveys at the micro-level (i.e., individual-level surveys), but because we are interested in the relationship between entrepreneurship and economic growth, we use the *country-level* measures of the APS data. These variables include opportunity-motivated entrepreneurship (OME), necessity-motivated entrepreneurship (NME), and a host of relevant control variables.

Both OME and NME come from total early-stage entrepreneurial activity (TEA), which is defined as the percentage of the adult population (18-64 years old) that is either actively involved in starting a new venture or is the owner/manager of a business that is less than 42 months old (Reynolds et al., 2005). OME reports the percentage of individuals who are actively involved in TEA, and who become an entrepreneur in order to take advantage of a business opportunity. NME instead reports the percentage of individuals who are actively involved in TEA, and who become an entrepreneur due to "no better choices for work" (Reynolds et al., 2004, p. 217). Because studies have found that OME and NME might have different effects on economic growth for different levels of country development (Acs et al., 2008; Ács & Varga, 2005; Sautet, 2013; Van Stel et al., 2005), we include both measures of entrepreneurship in our regression models. In total, 7.94 percent of individuals are classified as OME and 2.91 percent of individuals are classified as NME.

*3.2.3. Institutions: Entrepreneurial Framework Conditions*

GEM also collects data designed to measure the institutional conditions necessary for entrepreneurship known as the Entrepreneurial Framework Conditions:

> "Since its inception, GEM has proposed that entrepreneurship dynamics can be linked to conditions that enhance (or hinder) new business creation. In the GEM´s methodology these conditions are known as Entrepreneurial Framework Conditions (EFCs) (GEM, 2016)."

GEM collects data on EFCs in the National Expert Survey (NES), and we use the EFCs to construct our measure of institutions. First, we created a scale by combining 52 items associated with the nine constructs (measured on a 5-point Likert scale) that comprise GEM's EFCs: (1) Entrepreneurial Finance, (2) Government Policy, (3) Government Entrepreneurship Programs, (4) Entrepreneurship Education, (5) R&D Transfer, (6) Commercial & Legal Infrastructure, (7) Market Openness, (8) Physical Infrastructure, and (9) Cultural & Social Norms (GEM, 2016). Our scale has good internal consistency (Cronbach's Alpha = 0.96). Following recent work on the institutional drivers behind high-growth entrepreneurship (Krasniqi & Desai, 2016), we used Principal Component Analysis (PCA) and followed Kaiser's well-known criterion to retain factors with eigenvalues larger than one and to inspect the corresponding scree plot (Cattell, 1966). Our examination revealed a single underlying "latent" construct with all items loading positively and significantly to this factor. We report these constructs and the specific items used to create the EFCs in Table 2.

**[Insert Table 2]**

*3.2.4. Controls*

We also include several additional variables that have been shown to either affect entrepreneurship and institutions. We include a measure that captures the extent of opportunity recognition within the country because opportunity recognition is considered an important antecedent of entrepreneurial behavior (Kirzner, 1973, 1985; Klein, 2008; Schultz, 1975; Shane, 2000). Likewise, risk and uncertainty are inherent to the entrepreneurial process (Knight, 1921; McMullen & Shepherd, 2006), and some are deterred from entrepreneurship entry due to a fear of failure. Because studies include fear of failure as a potential deterrent to entrepreneurship (Boudreaux, Nikolaev, & Klein, 2018; Goltz, Buche, & Pathak, 2015; Wennberg, Pathak, & Autio, 2013; Xavier-Oliveira, Laplume, & Pathak, 2015), we include this variable as an additional control. We also include several additional measures that capture a society's perception of entrepreneurs—a variable that captures whether entrepreneurship is a desirable career choice, a variable that captures whether there is a high status for entrepreneurs, and a variable that captures whether media attention is given to entrepreneurs. Recent studies using GEM data have included these entrepreneurial perception measures for their potential to shape and influence entrepreneurship (Hechavarría, Terjesen, Stenholm, Brännback, & Lång, 2017). Lastly, we also include year and country dummies to control for geographical differences as well as differences over time (e.g., the great recession from 2007-2009). In addition, including these country and year dummies allows us to conduct a longitudinal analysis of the data, which changes the interpretation of our results to differences *within* countries over time rather than differences *between* countries, as is the case with cross-sectional data.

### 3.3. Model

We use a mixture model to test the hypothesis of two distinct groups of countries that each have a different effect of entrepreneurship on economic growth. The advantage of the mixture approach is that the assumption that all observations are drawn from a single underlying distribution is actually a testable hypothesis. Mixture models do not require prior information about whether different groups exist in the data—the estimation reveals whether there are distinct groups in the data (Caudill, Gropper, & Hartarska, 2009). Our analysis thus does not depend on sample selection decisions and criteria that might otherwise split observations arbitrarily into separate groups.

## 4. Results

We use mixture modeling to test the hypothesis that the effect of entrepreneurship on economic growth differs by the level of economic development. To assess the appropriateness of mixture modeling, we compare these results to Ordinary Least Squares (OLS) regression estimates for the full sample. We find evidence in support of the existence of two regimes. We observe that the standard errors ($\Sigma$) of the mixture model—0.021 for regime 1 and 0.031 for regime 2—are smaller than the standard error in the OLS model (0.07). This suggests the two regimes exist and that the mixture procedure is not simply "creamskimming" (Caudill et al., 2009, p. 663). We also report the mixing parameter that specifies the proportion of observations into each regime ($\Theta$) to ensure that the mixture model mixes observations appropriately. Roughly half of the observations behave according to the first regime and the other half behave according to the second regime. That is, 48.9 percent of countries have a positive relationship between OME and economic growth and 51.1 percent of countries have a negative relationship between NME and economic growth. We report these results in Table 3.

**[Insert Table 3]**

Our evidence supports the hypothesis that the effect of entrepreneurship on economic growth differs by the level of economic development. We observe that opportunity-motivated entrepreneurship (OME) is positively associated with economic growth for Regime 1 and necessity-motivated entrepreneurship (NME) is negatively associated with economic growth for Regime 2. The results from the mixture model, therefore, suggests that different regimes (i.e., groups) of countries have different effects of entrepreneurship on economic growth. Moreover, the standard error of each regime in the mixture model is smaller than OLS's standard error, which provides justification for the use of mixture modeling. We also observe that pro-market institutions—as measured by entrepreneurial framework conditions (EFCs)—are positively associated with economic growth in Regime 1 and have no effect on economic growth in Regime 2. We will now argue that each regime approximates the level of economic development for both developed countries (Regime 1) and developing countries (Regime 2).

While we have shown that different regimes have different effects of entrepreneurship on economic growth, we have only speculated that the regimes bifurcate the data by levels of economic development. If true, then we should be able to split our sample by different levels of economic development and find different effects of entrepreneurship on economic growth. Table 4 reports the findings from this exercise, where we compare the full sample to three subcategories of economic development based on quartiles of the income distribution: (i) low income, (ii) middle and upper income, and (iii) high income.[1] Our results in Table 4 report similar findings to the mixture model that we reported in Table 3. In low-income countries, NME is negatively associated with economic growth. In middle income, upper income, and high-income countries, however, we find a positive association between OME and economic growth. We also observe that EFCs are positively associated with economic growth for all countries except for low-income countries.

---

[1] Low income = quartile 1; middle and upper income = quartiles 2 and 3; high income = quartile 4

These findings, therefore, validate our findings in Table 3 and indicate that the regimes identified in the mixture model serve as proxies for the level of economic development.[2]

**[Insert Table 4]**

## 5. Discussion and Concluding Remarks

Our study examined the relationship between entrepreneurship and economic growth while hypothesizing that this relationship depends on the level of economic development. Specifically, we hypothesized that entrepreneurship positively contributes to economic growth in developed countries and negatively contributes to economic growth in developing countries. We based this hypothesis upon a reading of the literature on entrepreneurship, economic growth, and the role of pro-market institutions. Theoretical insights such as those from institutional economics and development economics helped us to understand how entrepreneurship functions as an underlying mechanism towards economic growth. In the Coleman bathtub model, for instance, pro-market institutions encourage entrepreneurial traits and characteristics, which in turn, contributes to higher rates of entrepreneurship. This ultimately has a positive effect on economic growth. This linkage is based on the idea that pro-market institutions help to encourage productive entrepreneurship or inhibit unproductive entrepreneurship (Baumol, 1990; Sobel, 2008) and supports recent developments in the literature (Bosma et al., 2018).

Our study also has important policy implications. While entrepreneurship leads to economic growth in some circumstances, we find that it has a limited effect on economic growth in developing countries. In fact, necessity-motivated entrepreneurship (NME) is negatively associated with economic growth in developing countries. As a result, policy makers might look to reduce the reliance on NME in developing countries. Based on our evidence, we would expect such policies to encourage higher rates of economic growth. Of course, policies designed to reduce NME must taken into consideration many different features (e.g., why do these individuals have no better options than to become entrepreneurs in the first place?). Clearly, there is no universal answer but policy makers in the developing world might look into alternative policies that can encourage opportunity-motivated entrepreneurship (OME) rather than NME. Although OME is not positively associated with economic growth in developing countries, this substitution from NME to OME might reduce the negative effect of entrepreneurship on economic growth in developing countries.

---

[2] One might question why we ought to bother with mixture modeling if the regressions based on the levels of economic development report similar findings. One reason is that we might not know *a priori* how the regimes differ. Even if we can hypothesize that, the regimes differ by some characteristic (e.g., economic development), mixture modeling is still beneficial because it removes some of the arbitrary decisions from empirical analysis such as, at which levels of economic development should we separate our samples? Results might be sensitive to the categories of classification, and we can circumvent this issue by allowing the statistical program to decide for us whether there are differences between groups or not.

Another policy implication is that, while the development of pro-market institutions might encourage economic growth in developed countries, it is unlikely to have an effect on economic growth in developing countries. Although we do not examine this source of heterogeneity, recent insights suggest that both formal and informal institutions are important for entrepreneurship and the effect of one type of institution on entrepreneurship might critically depend on the existence of the other (Krasniqi & Desai, 2016). Based on these insights, we can speculate that the formal institutions we use in our study might be less effective in developing countries because of a weak foundation of informal institutions (e.g., corruption). Of course, this is only speculation on our part. Future research, might consider why pro-market institutions do not have the same effect on economic growth in developing countries.

In sum, our study finds that entrepreneurship is important for economic growth, but it has different effects depending on the level of economic development. In middle and high-income countries, OME has a positive effect on economic growth. In low-income countries, however,

NME has a negative effect on economic growth. Therefore, the notion that entrepreneurship always encourages economic growth should be considered only in the appropriate context.

**References**

Acemoglu, D., Johnson, S., & Robinson, J. A. (2005). Chapter 6 Institutions as a Fundamental Cause of Long-Run Growth. In P. A. and S. N. Durlauf (Ed.), *Handbook of Economic Growth* (Vol. 1, Part A, pp. 385–472). Elsevier. Retrieved from http://www.sciencedirect.com/science/article/pii/S1574068405010063

Acs, Z. (2006). How is entrepreneurship good for economic growth? *Innovations*, *1*(1), 97–107. https://doi.org/10.1162/itgg.2006.1.1.97

Acs, Z., Åstebro, T., Audretsch, D., & Robinson, D. T. (2016). Public policy to promote entrepreneurship: a call to arms. *Small Business Economics*, *47*(1), 35–51. https://doi.org/10.1007/s11187-016-9712-2

Acs, Z., Braunerhjelm, P., Audretsch, D. B., & Carlsson, B. (2009). The knowledge spillover theory of entrepreneurship. *Small Business Economics*, *32*(1), 15–30. https://doi.org/10.1007/s11187-008-9157-3

Acs, Z., Desai, S., & Hessels, J. (2008). Entrepreneurship, economic development and institutions. *Small Business Economics*, *31*(3), 219–234. https://doi.org/10.1007/s11187-008-9135-9

Acs, Z., Estrin, S., Mickiewicz, T., & Szerb, L. (2017). *Institutions, Entrepreneurship and Growth: The Role of National Entrepreneurial Ecosystems* (SSRN Scholarly Paper No. ID 2912453). Rochester, NY: Social Science Research Network. Retrieved from https://papers.ssrn.com/abstract=2912453

Acs, Z. J., Arenius, P., Hay, M., & Minniti, M. (2004). Global entrepreneurship monitor.

Acs, Z. J., Audretsch, D. B., Braunerhjelm, P., & Carlsson, B. (2012). Growth and entrepreneurship. *Small Business Economics*, *39*(2), 289–300. https://doi.org/10.1007/s11187-010-9307-2

Acs, Z. J., Estrin, S., Mickiewicz, T., & Szerb, L. (2018). Entrepreneurship, institutional economics, and economic growth: an ecosystem perspective. *Small Business Economics*, *51*(2), 501–514. https://doi.org/10.1007/s11187-018-0013-9

Ács, Z. J., & Varga, A. (2005). Entrepreneurship, Agglomeration and Technological Change. *Small Business Economics*, *24*(3), 323–334. https://doi.org/10.1007/s11187-005-1998-4

Acs, Z., & Szerb, L. (2007). Entrepreneurship, Economic Growth and Public Policy. *Small Business Economics*, *28*(2–3), 109–122. https://doi.org/10.1007/s11187-006-9012-3

Alvarez, S. A., & Barney, J. B. (2014). Entrepreneurial opportunities and poverty alleviation. *Entrepreneurship Theory and Practice*, *38*(1), 159–184.

Anokhin, S., Grichnik, D., & Hisrich, R. D. (2008). The Journey from Novice to Serial Entrepreneurship in China and Germany: Are the drivers the same? *Managing Global Transitions*, *6*(2), 117.

Aquilina, M., Klump, R., & Pietrobelli, C. (2006). Factor Substitution, Average Firm Size and Economic Growth. *Small Business Economics*, *26*(3), 203–214. https://doi.org/10.1007/s11187-005-4715-4

Audretsch, D. B., & Keilbach, M. (2007). The theory of knowledge spillover entrepreneurship. *Journal of Management Studies*, *44*(7), 1242–1254.

Audretsch, D. B., Keilbach, M. C., & Lehmann, E. E. (2006). *Entrepreneurship and Economic Growth*. Oxford University Press, USA.

Autio, E. (2008). High-and low-aspiration entrepreneurship and economic growth in low-income economies (pp. 21–23). Presented at the UNU-WIDER Workshop on Entrepreneurship in Economic Development, Helsinki.

Baumol, W. J. (1986). Entrepreneurship and a century of growth. *Journal of Business Venturing*, *1*(2), 141–145.

Baumol, W. J. (1990). Entrepreneurship: Productive, Unproductive, and Destructive. *Journal of Political Economy*, *98*(5, Part 1), 893–921. https://doi.org/10.1086/261712

Baumol, W. J., & Strom, R. J. (2007). Entrepreneurship and economic growth. *Strategic Entrepreneurship Journal*, *1*(3–4), 233–237. https://doi.org/10.1002/sej.26

Bjørnskov, C., & Foss, N. (2013). How Strategic Entrepreneurship and The Institutional Context Drive Economic Growth. *Strategic Entrepreneurship Journal*, *7*(1), 50–69. https://doi.org/10.1002/sej.1148

Bjørnskov, C., & Foss, N. J. (2008). Economic freedom and entrepreneurial activity: Some cross-country evidence. *Public Choice*, *3*(134), 307–328. https://doi.org/10.1007/s11127-007-9229-y

Bjørnskov, C., & Foss, N. J. (2016). Institutions, Entrepreneurship, and Economic Growth: What Do We Know and What Do We Still Need to Know? *The Academy of Management Perspectives*, *30*(3), 292–315. https://doi.org/10.5465/amp.2015.0135

Bosma, N., Content, J., Sanders, M., & Stam, E. (2018). Institutions, entrepreneurship, and economic growth in Europe. *Small Business Economics*, *51*(2), 483–499. https://doi.org/10.1007/s11187-018-0012-x

Boudreaux, C. J. (2014). Jumping off of the Great Gatsby curve: how institutions facilitate entrepreneurship and intergenerational mobility. *Journal of Institutional Economics*, *10*(2), 231–255. https://doi.org/10.1017/S1744137414000034

Boudreaux, C. J., & Nikolaev, B. (2018). Capital is not enough: opportunity entrepreneurship and formal institutions. *Small Business Economics*, 1–30. https://doi.org/10.1007/s11187-018-0068-7

Boudreaux, C. J., Nikolaev, B., & Holcombe, R. (2018). Corruption and destructive entrepreneurship. *Small Business Economics*, *51*(1), 181–202. https://doi.org/10.1007/s11187-017-9927-x

Boudreaux, C. J., Nikolaev, B., & Klein, P. (2017). Entrepreneurial Traits, Institutions, and the Motivation to Engage in Entrepreneurship. *Academy of Management Proceedings*, *2017*(1), 16427. https://doi.org/10.5465/AMBPP.2017.33

Boudreaux, C. J., Nikolaev, B. N., & Klein, P. (2018). Socio-cognitive traits and entrepreneurship: The moderating role of economic institutions. *Journal of Business Venturing*.

Bowen, H. P., & Clercq, D. D. (2008). Institutional context and the allocation of entrepreneurial effort. *Journal of International Business Studies*, *39*(4), 747–767. https://doi.org/10.1057/palgrave.jibs.8400343

Bradley, S. W., & Klein, P. (2016). Institutions, Economic Freedom, and Entrepreneurship: The Contribution of Management Scholarship. *The Academy of Management Perspectives*, *30*(3), 211–221. https://doi.org/10.5465/amp.2013.0137

Braunerhjelm, P., Acs, Z. J., Audretsch, D. B., & Carlsson, B. (2010). The missing link: knowledge diffusion and entrepreneurship in endogenous growth. *Small Business Economics*, *34*(2), 105–125. https://doi.org/10.1007/s11187-009-9235-1

Braunerhjelm, P., Ding, D., & Thulin, P. (2018). The knowledge spillover theory of intrapreneurship. *Small Business Economics*, *51*(1), 1–30. https://doi.org/10.1007/s11187-017-9928-9

Bruton, G. D., Ketchen Jr, D. J., & Ireland, R. D. (2013). Entrepreneurship as a solution to poverty. *Journal of Business Venturing*, *28*(6), 683–689.

Carree, M., van Stel, A., Thurik, R., & Wennekers, S. (2002). Economic Development and Business Ownership: An Analysis Using Data of 23 OECD Countries in the Period 1976–1996. *Small Business Economics*, *19*(3), 271–290. https://doi.org/10.1023/A:1019604426387

Cattell, R. B. (1966). The Scree Test For The Number Of Factors. *Multivariate Behavioral Research*, *1*(2), 245–276. https://doi.org/10.1207/s15327906mbr0102_10

Caudill, S. B., Gropper, D. M., & Hartarska, V. (2009). Which Microfinance Institutions Are Becoming More Cost Effective with Time? Evidence from a Mixture Model. *Journal of Money, Credit and Banking*, *41*(4), 651–672. https://doi.org/10.1111/j.1538-4616.2009.00226.x

Coleman, J. S. (1990). *Foundations of Social Theory*. Cambridge, MA: Harvard University Press.

Court, J., & Maxwell, S. (2005). Policy entrepreneurship for poverty reduction: bridging research and policy in international development. *Journal of International Development: The Journal of the Development Studies Association*, *17*(6), 713–725.

Dawson, J. W. (1998). Institutions, Investment, and Growth: New Cross-Country and Panel Data Evidence. *Economic Inquiry*, *36*(4), 603–619. https://doi.org/10.1111/j.1465-7295.1998.tb01739.x

De Clercq, D., Lim, D., & Oh, C. (2013). Individual-Level Resources and New Business Activity: The Contingent Role of Institutional Context. *Entrepreneurship Theory and Practice*, *37*(2), 303–330. https://doi.org/10.1111/j.1540-6520.2011.00470.x

De Soto, H. (2000). *The Mystery of Capital: Why Capitalism Triumphs in the West and Fails Everywhere Else*. Basic Books.

Dejardin, M. (2000). Entrepreneurship and economic growth: An obvious conjunction. *The Institute for Development Strategies, Indiana University, Http://Econwpa. Wstl. Edu/Eps/Dev/Papers/0110/0110010. Pdf*.

Djankov, S., La Porta, R., Lopez-de-Silanes, F., & Shleifer, A. (2002). The Regulation of Entry. *The Quarterly Journal of Economics*, *117*(1), 1–37. https://doi.org/10.1162/003355302753399436

Dollar, D., & Kraay, A. (2003). Institutions, trade, and growth. *Journal of Monetary Economics*, *50*(1), 133–162. https://doi.org/10.1016/S0304-3932(02)00206-4

Douglas, E. J., & Shepherd, D. A. (2002). Self-employment as a Career Choice: Attitudes, Entrepreneurial Intentions, and Utility Maximization. *Entrepreneurship Theory and Practice*, *26*(3), 81–90.

GEM. (2016). *GEM*. Retrieved from http://www.gemconsortium.org

Goltz, S., Buche, M. W., & Pathak, S. (2015). Political Empowerment, Rule of Law, and Women's Entry into Entrepreneurship. *Journal of Small Business Management*, *53*(3), 605–626. https://doi.org/10.1111/jsbm.12177

Hechavarría, D. M., Terjesen, S. A., Stenholm, P., Brännback, M., & Lång, S. (2017). More than words: do gendered linguistic structures widen the gender gap in entrepreneurial activity? *Entrepreneurship Theory and Practice*.

Ho, Y.-P., & Wong, P.-K. (2007). Financing, Regulatory Costs and Entrepreneurial Propensity. *Small Business Economics*, *28*(2–3), 187–204. https://doi.org/10.1007/s11187-006-9015-0

Holcombe, R. G. (1998). Entrepreneurship and economic growth. *Quarterly Journal of Austrian Economics*, *1*(2), 45–62.

Islam, N. (1995). Growth Empirics: A Panel Data Approach. *The Quarterly Journal of Economics*, *110*(4), 1127–1170. https://doi.org/10.2307/2946651

Jorgenson, D. W. (2001). Information technology and the US economy. *American Economic Review*, *91*(1), 1–32.

Kim, P. H., Wennberg, K., & Croidieu, G. (2016). Untapped riches of meso-level applications in multilevel entrepreneurship mechanisms. *The Academy of Management Perspectives*, *30*(3), 273–291.

Kirzner, I. M. (1973). *Competition and Entrepreneurship* (New edition edition). Chicago: University Of Chicago Press.

Kirzner, I. M. (1985). *Discovery and the Capitalist Process*. Chicago: Univ of Chicago Pr.

Klapper, L., Laeven, L., & Rajan, R. (2006). Entry regulation as a barrier to entrepreneurship. *Journal of Financial Economics*, *82*(3), 591–629. https://doi.org/10.1016/j.jfineco.2005.09.006

Klein, P. G. (2008). Opportunity discovery, entrepreneurial action, and economic organization. *Strategic Entrepreneurship Journal*, *2*(3), 175–190. https://doi.org/10.1002/sej.50

Knight, F. (1921). *Risk, uncertainty and profit*. New York: Hart, Schaffner and Marx.

Krasniqi, B. A., & Desai, S. (2016). Institutional drivers of high-growth firms: country-level evidence from 26 transition economies. *Small Business Economics*, *47*(4), 1075–1094. https://doi.org/10.1007/s11187-016-9736-7

Lucas, R. E. (1988). On the mechanics of economic development. *Journal of Monetary Economics*, *22*(1), 3–42. https://doi.org/10.1016/0304-3932(88)90168-7

Mason, C., & Brown, R. (2013). Creating good public policy to support high-growth firms. *Small Business Economics*, *40*(2), 211–225. https://doi.org/10.1007/s11187-011-9369-9

McMullen, J. S., Bagby, D. R., & Palich, L. E. (2008). Economic Freedom and the Motivation to Engage in Entrepreneurial Action. *Entrepreneurship Theory and Practice*, *32*(5), 875–895. https://doi.org/10.1111/j.1540-6520.2008.00260.x

McMullen, J. S., & Shepherd, D. A. (2006). Entrepreneurial Action And The Role Of Uncertainty In The Theory Of The Entrepreneur. *Academy of Management Review*, *31*(1), 132–152. https://doi.org/10.5465/AMR.2006.19379628

Mueller, P. (2007). Exploiting Entrepreneurial Opportunities: The Impact of Entrepreneurship on Growth. *Small Business Economics*, *28*(4), 355–362. https://doi.org/10.1007/s11187-006-9035-9

Naudé, W. (2009). Entrepreneurship, developing countries, and development economics: new approaches and insights. *Small Business Economics*, *34*(1), 1. https://doi.org/10.1007/s11187-009-9198-2

Nikolaev, B., Boudreaux, C. J., & Palich, L. E. (2018). Cross-Country Determinants of Early Stage Necessity and Opportunity-Motivated Entrepreneurship: Accounting for Model Uncertainty. *Journal of Small Business Management*. https://doi.org/10.1111/jsbm.12400

North, D. C. (1990). *Institutions, Institutional Change and Economic Performance*. Cambridge University Press.

Nyström, K. (2008). The institutions of economic freedom and entrepreneurship: evidence from panel data. *Public Choice*, *136*(3–4), 269–282. https://doi.org/10.1007/s11127-008-9295-9

Porter, M. E. (1990). *The competitive advantage of nations*.

Reynolds, P., Bosma, N., Autio, E., Hunt, S., Bono, N. D., Servais, I., … Chin, N. (2005). Global Entrepreneurship Monitor: Data Collection Design and Implementation 1998–2003. *Small Business Economics*, *24*(3), 205–231. https://doi.org/10.1007/s11187-005-1980-1

Reynolds, P. D., Carter, N. M., Gartner, W. B., & Greene, P. G. (2004). The Prevalence of Nascent Entrepreneurs in the United States: Evidence from the Panel Study of Entrepreneurial Dynamics. *Small Business Economics*, *23*(4), 263–284. https://doi.org/10.1023/B:SBEJ.0000032046.59790.45

Romer, P. M. (1986). Increasing Returns and Long-Run Growth. *Journal of Political Economy*, *94*(5), 1002–1037. https://doi.org/10.1086/261420

Romer, P. M. (1990). Endogenous Technological Change. *Journal of Political Economy*, *98*(5, Part 2), S71–S102. https://doi.org/10.1086/261725

Sautet, F. (2013). Local and Systemic Entrepreneurship: Solving the Puzzle of Entrepreneurship and Economic Development. *Entrepreneurship Theory and Practice*, *37*(2), 387–402. https://doi.org/10.1111/j.1540-6520.2011.00469.x

Schultz, T. W. (1975). The Value of the Ability to Deal with Disequilibria. *Journal of Economic Literature*, *13*(3), 827–846.

Schumpeter, J. A. (1934). *The Theory of Economic Development: An Inquiry Into Profits, Capital, Credit, Interest, and the Business Cycle*. Transaction Publishers.

Scott, W. R. (1995). Institutions and organizations. Foundations for organizational science. *London: A Sage Publication Series*.

Shane, S. (2000). Prior Knowledge and the Discovery of Entrepreneurial Opportunities. *Organization Science*, *11*(4), 448–469. https://doi.org/10.1287/orsc.11.4.448.14602

Shane, S. (2009). Why encouraging more people to become entrepreneurs is bad public policy. *Small Business Economics*, *33*(2), 141–149. https://doi.org/10.1007/s11187-009-9215-5

Sobel, R. S. (2008). Testing Baumol: Institutional quality and the productivity of entrepreneurship. *Journal of Business Venturing*, *23*(6), 641–655. https://doi.org/10.1016/j.jbusvent.2008.01.004

Sternberg, R., & Wennekers, S. (2005). Determinants and Effects of New Business Creation Using Global Entrepreneurship Monitor Data. *Small Business Economics*, *24*(3), 193–203.

Urbano, D., & Alvarez, C. (2014). Institutional dimensions and entrepreneurial activity: an international study. *Small Business Economics*, *42*(4), 703–716. https://doi.org/10.1007/s11187-013-9523-7

Urbano, D., Aparicio, S., & Audretsch, D. (2018). Twenty-five years of research on institutions, entrepreneurship, and economic growth: what has been learned? *Small Business Economics*. https://doi.org/10.1007/s11187-018-0038-0

Van Stel, A., Carree, M., & Thurik, R. (2005). The Effect of Entrepreneurial Activity on National Economic Growth. *Small Business Economics*, *24*(3), 311–321. https://doi.org/10.1007/s11187-005-1996-6

Van Stel, A., Storey, D. J., & Thurik, A. R. (2007). The Effect of Business Regulations on Nascent and Young Business Entrepreneurship. *Small Business Economics*, *28*(2–3), 171–186. https://doi.org/10.1007/s11187-006-9014-1

Wennberg, K., Pathak, S., & Autio, E. (2013). How culture moulds the effects of self-efficacy and fear of failure on entrepreneurship. *Entrepreneurship & Regional Development*, *25*(9–10), 756–780. https://doi.org/10.1080/08985626.2013.862975

Wennekers, S., & Thurik, R. (1999). Linking Entrepreneurship and Economic Growth. *Small Business Economics*, *13*(1), 27–56. https://doi.org/10.1023/A:1008063200484

Wiklund, J., Davidsson, P., & Delmar, F. (2003). What Do They Think and Feel about Growth? An Expectancy–Value Approach to Small Business Managers' Attitudes Toward Growth. *Entrepreneurship Theory and Practice*, *27*(3), 247–270. https://doi.org/10.1111/1540-8520.00014

Xavier-Oliveira, E., Laplume, A. O., & Pathak, S. (2015). What motivates entrepreneurial entry under economic inequality? The role of human and financial capital. *Human Relations*, *68*(7), 1183–1207. https://doi.org/10.1177/0018726715578200

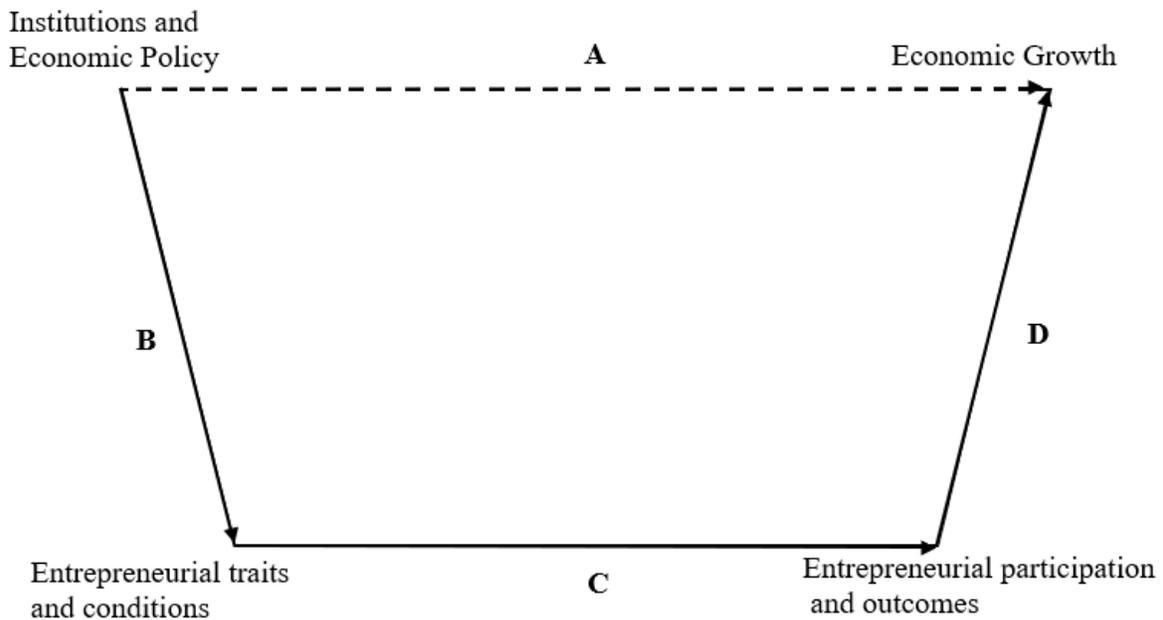

**Figure 1.** Using Coleman's Bathtub Model to Explain How Entrepreneurship Affects Economic Growth

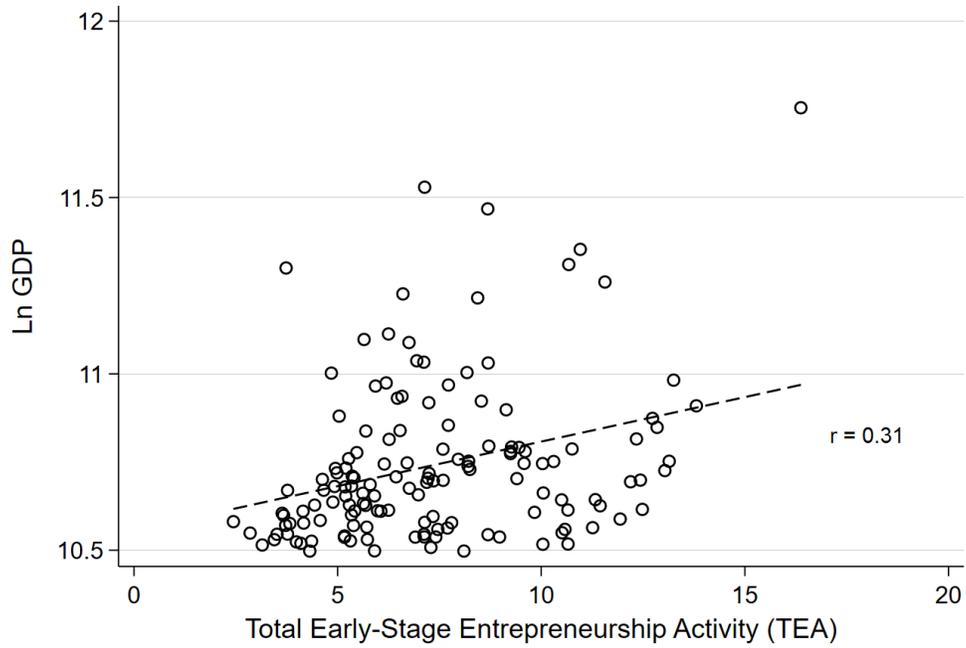

**Figure 2.** The Relationship between Entrepreneurship and Economic Growth for High-Income Countries

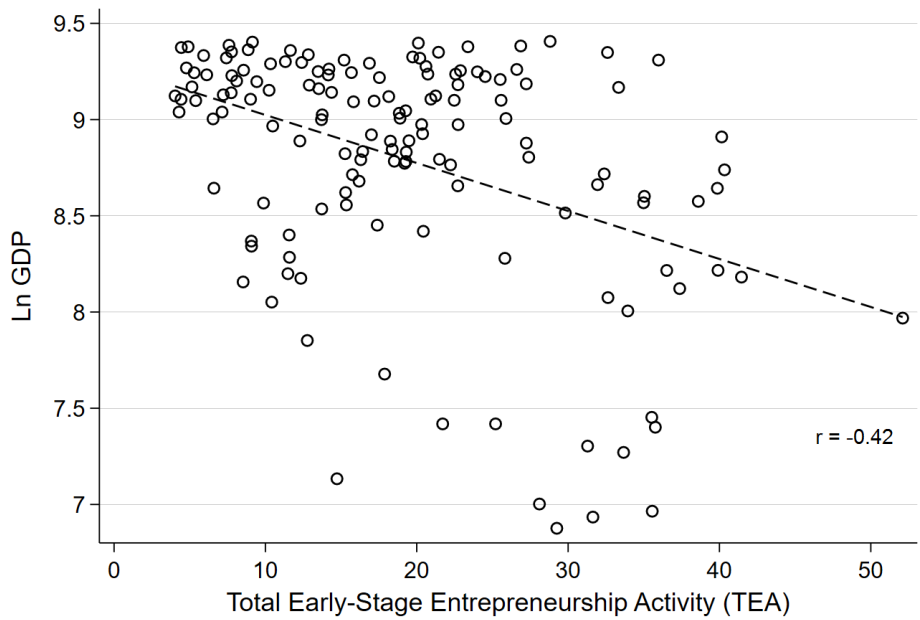

**Figure 3.** The Relationship between Entrepreneurship and Economic Growth for Low-Income Countries

**Table 1.** Summary statistics and correlation matrix

| | Mean | SD | | [1] | [2] | [3] | [4] | [5] | [6] | [7] | [8] | [9] | [10] | [11] |
|---|---|---|---|---|---|---|---|---|---|---|---|---|---|---|
| GDP per capita (PPP) $1000, | 25.8 | 17.3 | [1] | 1 | | | | | | | | | | |
| TEA | | | | | | | | | | | | | | |
|   Opportunity-motivated entrepreneurship (%) | 7.94 | 5.44 | [2] | -0.36* | 1 | | | | | | | | | |
|   Necessity-motivated entrepreneurship (%) | 2.91 | 2.95 | [3] | -0.57* | 0.77* | 1 | | | | | | | | |
| Entrepreneurial framework conditions (EFCs) | 2.76 | 0.29 | [4] | 0.62* | -0.22* | -0.45* | 1 | | | | | | | |
| Know other entrepreneurs (%) | 40.45 | 12.17 | [5] | -0.41* | 0.49* | 0.46* | -0.13* | 1 | | | | | | |
| Opportunity recognition (%) | 41.13 | 16.39 | [6] | -0.24* | 0.64* | 0.50* | -0.07 | 0.59* | 1 | | | | | |
| Entrepreneurial-self efficacy (%) | 50.37 | 14.44 | [7] | -0.49* | 0.66* | 0.68* | -0.39* | 0.54* | 0.61* | 1 | | | | |
| Fear of failure (%) | 37.46 | 9.62 | [8] | 0.13* | -0.25* | -0.22* | -0.05 | -0.26* | -0.36* | -0.29* | 1 | | | |
| Entrepreneurship is a desirable career choice (%) | 65.16 | 13.34 | [9] | -0.51* | 0.44* | 0.54* | -0.40* | 0.28* | 0.41* | 0.62* | -0.13* | 1 | | |
| High status for entrepreneurs (%) | 70.06 | 10.62 | [10] | -0.06 | 0.22* | 0.24* | 0.01 | 0.28* | 0.42* | 0.29* | -0.05 | 0.32* | 1 | |
| Media attention for entrepreneurs (%) | 60.34 | 15.03 | [11] | -0.15* | 0.46* | 0.39* | 0.07 | 0.39* | 0.51* | 0.35* | -0.28* | 0.39* | 0.41* | 1 |

*Note* - * $p < 0.05$.

**Table 2.** Rotated factor solution for Entrepreneurial Framework Conditions (EFCs)

| Items | Factor 1 |
| --- | --- |
| | EFC |
| 1. Entrepreneurial Finance | 0.808 |
| 2. Government Policy | 0.812 |
| 3. Government Entrepreneurship Programs | 0.799 |
| 4. Entrepreneurship Education | 0.679 |
| 5. R&D Transfer | 0.869 |
| 6. Commercial and Legal Infrastructure | 0.740 |
| 7. Entry Regulation | 0.693 |
| 8. Physical infrastructure | 0.719 |
| 9. Cultural and Social Norms | 0.617 |
| Cumulative variance explained | 56.61% |

Extraction method: principal component analysis.

Rotation method: varimax with Kaiser normalization

**Table 3.** Mixture model results for the effect of entrepreneurship on economic growth

|  | OLS | Mixture Model | |
| --- | --- | --- | --- |
|  | Full Sample | Regime 1 | Regime 2 |
| TEA | (1) | (2) | (3) |
|   Necessity-Motivated Entrepreneurship (%) | -0.011*** | -0.007 | -0.006*** |
|  | (0.004) | (0.005) | (0.002) |
|   Opportunity-Motivated Entrepreneurship (%) | 0.006*** | 0.005*** | 0.0003 |
|  | (0.002) | (0.002) | (0.001) |
| Entrepreneurial Framework Conditions (EFCs) | 0.036*** | 0.048*** | 0.013 |
|  | (0.008) | (0.014) | (0.026) |
| Know other entrepreneurs (%) | -0.000 | 0.0001 | 0.002** |
|  | (0.001) | (0.0005) | (0.0007) |
| Opportunity recognition (%) | 0.001* | 0.0003 | -0.001** |
|  | (0.001) | (0.0004) | (0.0005) |
| Entrepreneurial self-efficacy (%) | -0.000 | -0.0007 | 0.001* |
|  | (0.001) | (0.0006) | (0.001) |
| Fear of failure (%) | -0.003*** | -0.002*** | 0.001 |
|  | (0.001) | (0.0005) | (0.001) |
| Entrepreneurship is a desirable choice (%) | -0.001 | 0.004*** | -0.003*** |
|  | (0.001) | (0.001) | (0.001) |
| High status for entrepreneurs (%) | 0.001 | -0.001 | 0.002** |
|  | (0.001) | (0.001) | (0.001) |
| Media attention for entrepreneurs (%) | 0.000 | -0.001*** | 0.0003 |
|  | (0.001) | (0.0004) | (0.0005) |
| Country dummies? | Yes | Yes | Yes |
| Year dummies? | Yes | Yes | Yes |
| $\Sigma$ | 0.07 | 0.021 | 0.031 |
| $\Theta$ | – | 0.489 | 0.511 |
| $R^2$ | 0.825 | – | – |
| $F$ | 75.6*** | – | – |

*Note* – The dependent variable is Ln (GDP). N = 441 observations. $\Sigma$ reports the model's standard error. $\Theta$ reports the mixing parameter that specifies the proportion of observations into each regime. $R^2$ and $F$ are goodness-of-fit measures. Standard errors reported in parentheses (two-tailed test):
* p<0.10
** p<0.05
*** p<0.01

**Table 4.** Regression results for the effect of entrepreneurship on economic growth

|  | Full Sample | Low Income | Middle and Upper Income | High Income |
|---|---|---|---|---|
|  | (1) | (2) | (3) | (4) |
| TEA |  |  |  |  |
|   Opportunity-Motivated Entrepreneurship (%) | 0.020*** | 0.001 | 0.024*** | 0.016** |
|  | (0.003) | (0.006) | (0.004) | (0.007) |
|   Necessity-Motivated Entrepreneurship (%) | -0.018*** | -0.020** | 0.016 | 0.052* |
|  | (0.007) | (0.009) | (0.010) | (0.030) |
| Entrepreneurial Framework Conditions (EFCs) | 0.247*** | 0.080 | 0.253*** | 0.222*** |
|  | (0.046) | (0.116) | (0.059) | (0.067) |
| Know other entrepreneurs (%) | -0.010*** | -0.011*** | -0.006*** | -0.014*** |
|  | (0.001) | (0.003) | (0.001) | (0.002) |
| Opportunity recognition (%) | 0.002* | 0.003 | -0.000 | 0.001 |
|  | (0.001) | (0.003) | (0.001) | (0.001) |
| Entrepreneurial self-efficacy (%) | 0.005*** | 0.002 | 0.007*** | -0.004 |
|  | (0.002) | (0.004) | (0.002) | (0.003) |
| Fear of failure (%) | 0.003* | 0.009** | 0.002 | -0.002 |
|  | (0.001) | (0.004) | (0.002) | (0.002) |
| Entrepreneurship is a desirable choice (%) | -0.007*** | -0.014*** | -0.005*** | -0.007*** |
|  | (0.001) | (0.004) | (0.002) | (0.002) |
| High status for entrepreneurs (%) | 0.007*** | 0.015*** | 0.001 | 0.011*** |
|  | (0.002) | (0.003) | (0.002) | (0.002) |
| Media attention for entrepreneurs (%) | 0.000 | 0.004 | 0.001 | 0.004* |
|  | (0.001) | (0.003) | (0.001) | (0.002) |
| Country dummies? | Yes | Yes | Yes | Yes |
| Year dummies? | Yes | Yes | Yes | Yes |
| Intercept | 9.073*** | 8.421*** | 9.180*** | 10.011*** |
|  | (0.169) | (0.446) | (0.184) | (0.350) |
| Number of observations | 441 | 112 | 217 | 112 |
| Number of groups (countries) | 83 | 35 | 46 | 23 |
| $R^2$ | 0.39 | 0.43 | 0.54 | 0.65 |
| $F$ | 22.58*** | 5.11*** | 18.79*** | 14.84*** |

*Note* - The dependent variable is Ln (GDP). Country and year fixed effects included in all models. Standard errors reported in parentheses (two-tailed test):
* $p<0.10$
** $p<0.05$
*** $p<0.01$

# APPENDIX

**Table A1**
Measures and descriptive statistics.

| Variables | Measures | Mean | SD |
|---|---|---|---|
| OME | Opportunity-motivated entrepreneurship (OME) is the percentage of all respondents (18-64) who are involved in total early-stage entrepreneurial activity (TEA) to take advantage of an opportunity. | 7.94 | 5.44 |
| NME | Necessity-motivated entrepreneurship (NME) is the percentage of all respondents (18-64) who are involved in total early-stage entrepreneurial activity (TEA) because there are no better opportunities. | 2.91 | 2.95 |
| Opportunity recognition | Percentage of all respondents (18-64) who think that in the next 6 months there will be good opportunities for starting a business in the area where they live | 41.13 | 16.39 |
| Fear of failure | Percentage of all respondents (18-64) who said fear of failure would prevent them from starting a new business | 37.46 | 9.62 |
| Entrepreneurial self-efficacy | Percentage of all respondents (18-64) who say they have the knowledge, skill, and experience required to start a new business. | 50.37 | 14.44 |
| Know other entrepreneurs | Percentage of all respondents (18-64) who know someone personally who started a business in the past 2 years. | 40.45 | 12.17 |
| Entrepreneurship is a desirable career choice | Percentage of all respondents (18-64) who believe that in their country, most people consider starting a new business a desirable career choice | 65.16 | 13.34 |
| High status for entrepreneurs | Percentage of all respondents (18-64) who believe that in their country, those successful at starting a new business have a high level of status and respect. | 70.06 | 10.62 |
| Media attention for entrepreneurship | Percentage of all respondents (18-64) who believe that in their country, you will often see stories in the public media about successful new businesses. | 60.34 | 15.03 |
| Gross domestic product (GDP) | GDP per capita measured in inflation-adjusted US dollars (PPP international conversion). We transform this measure using the natural logarithm. | 25,826 | 17,343 |
| Entrepreneurial Framework Conditions (EFCs) | The average of 52 items associated with the nine constructs (measured on a 5-point Likert scale) that comprise GEM's EFCs: (1) Entrepreneurial Finance, (2) Government Policy, (3) Government Entrepreneurship Programs, (4) Entrepreneurship Education, (5) R&D Transfer, (6) Commercial & Legal Infrastructure, (7) Market Openness, (8) Physical Infrastructure, and (9) Cultural & Social Norms. [Cronbach's Alpha = 0.96]. | 2.80 | 0.29 |